\documentclass{aastex7}
\usepackage{graphicx}
\usepackage{subcaption}
\usepackage{caption}
\usepackage{tikz}
\usepackage{amsmath}  
\usepackage{xcolor} 
\usepackage[T1]{fontenc}

\shorttitle{Asteroids Pebble accretion}

\begin{document}

\title{Testing the Icy Pebble Accretion Hypothesis with Primordial Main Belt Asteroids}

\correspondingauthor{Jinfei Yu}
\email{jinfei-yu@g.ecc.u-tokyo.ac.jp}

\author[0000-0002-8793-3107]{Jinfei Yu}
\affiliation{Department of General Systems Studies, Graduate School of Arts and Sciences, The University of Tokyo \\
3-8-1 Komaba, Meguro-ku, Tokyo 153-8902, Japan}
\email{jinfei-yu@g.ecc.u-tokyo.ac.jp}

\author[0000-0003-1965-1586]{Hiroyuki Kurokawa}
\affiliation{Department of General Systems Studies, Graduate School of Arts and Sciences, The University of Tokyo \\
3-8-1 Komaba, Meguro-ku, Tokyo 153-8902, Japan}
\affiliation{Department of Earth and Planetary Science, Graduate School of Science, The University of Tokyo, \\
7-3-1 Hongo, Bunkyo-ku, Tokyo 113-0033, Japan}
\email{hirokurokawa@g.ecc.u-tokyo.ac.jp}

\author[0000-0002-6602-7113]{Tetsuo Taki}
\affiliation{Department of General Systems Studies, Graduate School of Arts and Sciences, The University of Tokyo \\
3-8-1 Komaba, Meguro-ku, Tokyo 153-8902, Japan}
\email{takitetsuo@g.ecc.u-tokyo.ac.jp}

\begin{abstract}

Large main-belt asteroids (diameter $D \gtrsim 120\ \mathrm{km}$) exhibit a surface composition gradient as a function of heliocentric distance, ranging from anhydrous bodies to those rich in hydrated and, possibly, ammoniated materials. Their primordial nature holds key clues to the evolution of the Solar System. It has been suggested that the volatile-rich bodies formed in the outer Solar System and were implanted into the main belt. Alternatively, volatiles may have been delivered via inward-drifting icy pebbles in the protosolar disk. Here, we examine whether in-situ formed rocky embryos can acquire volatiles through pebble accretion as the snowline migrated inward. With the turbulence strength of the disk, radial pebble flux, and the dimensionless stopping time of pebbles scaled with the Keplerian frequency ($\mathrm{St}$) as parameters, we calculate the growth of large asteroids. The results are then compared with mass and compositional constraints based on asteroid observations. We find that a moderate pebble flux ($\lesssim18~M_\oplus / \text{Myr}$) is required to enable volatile delivery while prevent the largest asteroids from becoming more massive than Ceres. Water accretion is feasible with $\mathrm{St} \sim 10^{-3}$ ($\sim 1$ mm). However, only the largest asteroids ($D \gtrsim 200$ km) can accumulate sufficient ammonia under such conditions. For most asteroids with $D \simeq 100$--$200\ \mathrm{km}$, ammonia ice accretion requires $\mathrm{St} \sim 10^{-4}$ ($\sim 100\,\mu$m). Such small particle sizes may pose both theoretical and observational challenges. Thus, we propose that the intermediate-sized, potentially ammonia-bearing asteroids serve as a record of the Solar System’s dynamic evolution.

\end{abstract}



\section{Introduction} \label{sec:introduction}

Evidence from astronomy and cosmochemistry suggests that the Solar System and other planetary systems originate from the accretion of gas and dust within a protoplanetary disk \citep{Lauretta2006,armitage2020astrophysics,annurev:/content/journals/10.1146/annurev-astro-022823-040820}. Two major mechanisms have been proposed for planet formation: planetesimal accretion and pebble accretion. Both theories begin with the formation of planetesimals (typically $\sim 100$ km in size) through, for instance, gravitational collapse once the local midplane dust density becomes high enough \citep{johansen2007rapid}. The planetesimal accretion model describes a growth process through mutual collisions \citep{safronov1972evolution}, with planetesimals forming planetary embryos with masses comparable to the Moon or Mars. After the dissipation of the gas disk, these embryos experience giant impacts and further accrete remaining planetesimals, forming planets over several million years \citep[for a comprehensive review, see][]{morbidelli2012building}. Another proposed mechanism is the pebble accretion model, which suggests that planets grow efficiently by accreting millimeter- to centimeter-sized pebbles drifting inward from the outer disk \citep{ormel2010effect, johansen2010prograde,lambrechts2012rapid,lambrechts2014forming}. This mechanism provides a plausible explanation for the rapid formation of giant planet cores within the lifetime of the protoplanetary disk \citep{johansen2017forming}. However, its role in shaping the Solar System remains highly debated \citep[e.g.,][]{morbidelli2025did,johansen2024comment}.

These planet formation mechanisms also play a key role in shaping planetary volatile species \citep[see][for a review]{broadley2022volatile}, such as water (H$_2$O), carbon dioxide (CO$_2$), and ammonia (NH$_3$), which are crucial ingredients for the formation and evolution of habitable planets such as Earth and for supporting life. The origin of volatiles in the inner Solar System remains an open question \citep{obrien2018delivery}, with competing theories suggesting that they were implanted via giant planet formation and migration \citep{walsh2011low} or transported inward through pebble drift as the snowline migrated within the protoplanetary disk \citep{sato2016,ida2019A&A...624A..28I}. Unfortunately, the various processes that terrestrial planets undergo, such as differentiation and degassing \citep{broadley2022volatile}, have erased much of the evidence left behind by early planet formation and volatile delivery. It is nearly impossible to reconstruct the initial conditions and evolutionary processes of planet formation solely by studying planets. Instead, small Solar System bodies (e.g., asteroids and comets) and meteorites retain primitive records with minimal alteration \citep{Lauretta2006,michel2015asteroids}, providing a crucial window into the formation of planets and the transport of volatiles in the early Solar System. 

The main asteroid belt, located between the terrestrial planets and the outer gas giants, represents a key transition region in the Solar System. This zone contains millions of irregularly shaped bodies larger than 1 km in diameter ($D$), composed of rock, ice, and metals, with a total mass of approximately 4\% that of the Moon \citep{Martin2012}. Meteorites sampled on Earth are believed to originate from the main-belt asteroids (hereafter MBAs) \citep{binzel1996spectral,bottke2002debiased, morbidelli1998orbital}, making MBAs a crucial population for understanding early Solar System evolution.

Over the past half-century, advances in the discovery and characterization of asteroids have revealed a systematic variation in the distribution of compositional types with heliocentric distance in the main belt \citep{gradie1982compositional,demeo2014solar}, the so-called S–C dichotomy—i.e., a radial split between stony (silicate-rich, ‘S’) asteroids that dominate the inner main belt and carbonaceous (‘C’) asteroids that prevail in the mid- to outer main belt. Specifically, visible and near-infrared (VNIR) spectroscopic taxonomy \citep[based on the SMASSII taxonomy,][]{bus2002phase} grouped the asteroid population into distinct and separate groups. S-complex asteroids (including S, A, R, K, L, and V types) dominate the inner main belt and are primarily composed of anhydrous silicates and exhibit a 1 $\mu$m absorption and are closely associated with ordinary chondrites \citep{nakamura2011itokawa}. C-complex asteroids (such as C-, D-, G-, F-, and B-types) are most commonly found in the mid-to-outer main belt (2.5--4.0 au); they contain volatile-rich materials such as phyllosilicates, exhibit a 0.7 $\mu$m absorption band and a reddish spectral slope, and are linked to carbonaceous chondrites, hereafter CCs \citep{demeo2013taxonomic,beck2021water}. Among the known MBAs, this compositional trend is most pronounced in the $\sim$140 large MBAs \citep[D $\gtrsim$ 120 km;][]{demeo2014solar,demeo2015compositional}, which are believed to be primordial with their properties likely determined during the first accretion epoch rather than during the collisional evolution \citep{bottke2005linking}. Thus, the large MBAs are relics of the early Solar System. Among these, Ceres stands out as the largest MBA (dwarf planet), with a diameter of $\sim950$ km and a bulk density of $\sim$2.1 g\,cm$^{-3}$, suggesting a water-ice fraction of $\sim$30\% \citep{prettyman2017extensive}. The low average impact velocities in the main belt ($\sim$5--6 km\,s$^{-1}$) make catastrophic disruption unlikely for such large bodies \citep{bottke1994velocity,o2011origin}, and the absence of large impact basins ($\gtrsim$280 km) on its surface \citep{Marchi2016Ceres} further supports the idea that Ceres has preserved its primordial structure and volatile inventory from the accretion period.

While the S–C dichotomy is primarily defined by VNIR spectra in 0.4--2.5 $\mu$m, observations at longer wavelengths (i.e., near-infrared; NIR) revealed a notable diversity in C-complex asteroids. Diverse absorption features can be found in the 3 $\mu$m region, where diagnostic features of OH/H$_2$O ($\sim$2.7--2.95 $\mu$m, hydrous minerals), water ice ($\sim$3.0 $\mu$m), N–H ($\sim$3.1 $\mu$m, ammoniated materials), C–H ($\sim$3.4 $\mu$m, aliphatic and aromatic organics), and C–O ($\sim$3.4 $\mu$m, carbonates) are present, collectively referred to as the 3 $\mu$m band \citep{beck2021water,yu2024near}. Prior studies in 3 $\mu$m band observations have revealed hints of diversity beyond the hydrated minerals typical of carbonaceous chondrites \citep{takir2012outer,usui2019akari}. This includes the presence of carbonates \citep{de2016bright,kaplan2020bright}, water ice \citep{campins2010water}, and even ammonia-bearing compounds \citep{1992Sci...255.1551K,Hsieh2006,de2015ammoniated} and carbon dioxide ice \citep{Hsieh2006,wong2024jwst}. Notably, ammonia-related absorption features were once thought to exist only on the largest asteroid, Ceres. However, recent observational advances have revealed similar features at $\sim3.1~\mu$m on other MBAs, suggesting that ammonia may be more widespread than previously assumed \citep{kurokawa2022distant,rivkin2022nature,takir2023late}. Moreover, the 3 $\mu$m band of C-complex asteroids appears to exhibit a dichotomy correlated with spatial distribution \citep{rivkin2022nature}; asteroids closer to the Sun predominantly show pronounced absorption associated with hydrated minerals similar to those in CCs (Sharp-types), while those farther from the Sun exhibit rounded spectral features (Not-Sharp-types) of ammonia, water ice, and carbon dioxide \citep{wong2024jwst}. Among observed large asteroids, these bands appear to exhibit a globally uniform distribution \citep{ammannito2016distribution,rivkin2022nature}, and their spectral shapes differ from the $\sim$3~$\mu$m features typically associated with pure surface ice or cometary material. Instead, they are more consistent with the presence of phyllosilicates formed through aqueous alteration, pointing to an endogenous origin rather than delivery via meteoritic or cometary infall. Moreover, exposed surface ice in the main belt would sublimate on timescales $\ll$1~Myr, making the long-term preservation of a cometary veneer unlikely \citep{prialnik2009long}.

The hypothesis of orbital migration provides a possible explanation for the observed compositional distribution of the asteroid belt \citep{takir2023late}. Such a mechanism has been widely invoked to explain the S-C dichotomy. Originally, these two types of bodies probably formed at a much wider range of heliocentric distances, as suggested by the distinct chemical and isotopic compositions \citep{burkhardt2011molybdenum,kruijer2017age} of meteorites connected with them. These anomalies are produced by the uneven distribution of isotopically anomalous presolar components and vary with heliocentric distance \citep{warren2011stable}, thus distinguishing between 'non-carbonaceous' and 'carbonaceous' meteorite reservoirs (i.e., NC-CC dichotomy). Gravitational scattering eventually led to the current distribution \citep{walsh2011low, Tsiganis2005}. 

Alternatively, ices may have migrated inward through pebble drift, a process where small, icy pebbles drift inward from the outer Solar System, and accreted onto forming planetesimals in the main belt \citep{de2015ammoniated,nara2019ESS.....431719N}. It is supported by recent models of protoplanetary disk thermal evolution, which suggest that magnetorotational instability, a key driver of disk turbulence, is suppressed in the midplane due to low ionization, resulting in lower viscous heating and a colder disk \citep{mori2021evolution}. As a result, volatile snowlines including those of water, ammonia, and carbon dioxide could have reached the main belt, enabling the later delivery of volatiles to the inner Solar System.

The overarching goal of this study is to evaluate how volatile transport and accretion in the early Solar System shaped the composition of small bodies and to use these insights to constrain planetary formation models. Specifically, we examine whether the icy pebble accretion hypothesis can account for the formation of volatile-rich asteroids in the main belt, as previously proposed for the origin of Ceres' ammoniated material \citep{de2015ammoniated,nara2019ESS.....431719N}, and whether their compositional diversity aligns with the predicted outcomes of different Solar System formation scenarios. Using a simplified disk model to describe the radial inward drift of icy pebbles and an analytic framework to calculate the pebble accretion rate, we aim to test whether these mechanisms can effectively explain masses and compositions of MBAs and provide insight into early Solar System formation. Section \ref{sec:methods} introduces our model. Section \ref{sec:results} presents our results. Section \ref{sec:discussion} discusses the implications for the origins of MBAs and addresses the limitations and validity of our model. Section \ref{sec:summary} summarizes this study.

\section{Methods}
\label{sec:methods}

\begin{figure}[htbp]
    \centering
    \includegraphics[width=0.7\textwidth]{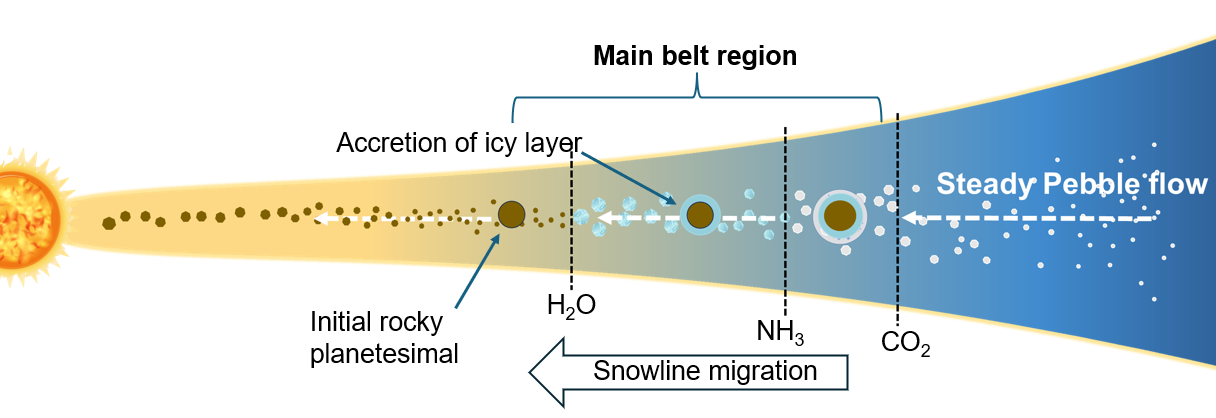} 
    \caption{Schematic illustration of the modeled scenario: inward migration of the snowlines (water, ammonia, and carbon dioxide) and the accretion of icy pebbles drifting inward through the protoplanetary disk onto rocky planetesimals formed in the main belt.}
    \label{fig:model}
\end{figure}

To understand the conditions under which asteroids form through pebble accretion and to estimate the amount of material delivered to their parent bodies (planetesimals), we adopted a simplified protoplanetary disk model to describe the radial drift of icy pebbles and used an analytic method to calculate the pebble accretion rate (see Figure~\ref{fig:model} for a schematic illustration of our model). Following \cite{sato2016}, we considered a static gas disk surrounding a solar-mass star without incorporating disk evolution except for temperature. This simplification allowed us to focus on the key parameters governing pebble accretion while avoiding uncertainties associated with disk dissipation timescales and gas-driven migration. Previous studies have employed specific models for dust growth, where the particle size could be derived rather than assumed \citep[e.g.,][]{birnstiel2012modeling,sato2016}. However, dust growth is subject to multiple uncertainties. This include fragmentation, bouncing barriers, as well as ice sublimation, condensation, and sintering, related to disk thermal evolution, which significantly affected the final particle size distribution \citep[e.g.,][]{zsom2010outcome,okuzumi2012rapid,testi2014dust,okuzumi2022global}. Given these uncertainties, we adopted a parameterized approach to treat the size (more specifically, the Stokes number $\mathrm{St}$) and radial mass flux of drifting pebbles $\dot{M}_\mathrm{d}$ as free parameters. Moreover, we treated the turbulence strength $\alpha$ of the disk as another parameter, given its uncertainty \citep[e.g.,][]{Lesur2023ASPC..534..465L}. For planetesimals with given masses (radii) and orbits, the accretion efficiency of pebbles is calculated using analytical expressions. This allows us to systematically explore how variations in these parameters influence pebble accretion efficiency and volatile transport.

\subsection{Disk Model}
\label{disk}
\subsubsection{Gas Profile}

The radial distribution of the gas surface density, $\Sigma_\mathrm{g}$, is taken from the minimum mass solar nebula (MMSN) model of \cite{1981PThPS..70...35H}, as given by, 
\begin{equation}
    \Sigma_\mathrm{g}(r) = 1700 \left( \frac{r}{1\,\mathrm{au}} \right)^{-3/2}\, \mathrm{g/cm^2},
\label{eq1}
\end{equation}
where $r$ is the distance from the central star. The gas disk is assumed to be isothermal and hydrostatic in the vertical direction. Under these conditions, the gas density at the midplane can be expressed as
\(
\rho_\mathrm{g} = {\Sigma_\mathrm{g}}/({\sqrt{2\pi} h_\mathrm{g}}),
\)
where \( h_g = {c_\mathrm{s}}/{\Omega_\mathrm{K}} \) represents the gas scale height, \( c_\mathrm{s} = \sqrt{{k_\mathrm{B} T}/{m_\mathrm{g}}} \) denotes the isothermal sound speed, \( k_\mathrm{B} \) is the Boltzmann constant, \( m_\mathrm{g} \) is the mean molecular mass of disk gas (assumed to be 2.34 amu). The Keplerian angular frequency is \(\Omega_K=\sqrt{G M_\odot/r^3}\), \( G \) is the gravitational constant, and $\Omega_K$ can be defined as:
\begin{equation}
    \Omega_\mathrm{K} = 2.0 \times 10^{-7}  \left(\frac{r}{1\,\mathrm{au}}\right)^{-3/2} \, \mathrm{s^{-1}}.
\end{equation}

\begin{figure}[htbp]
    \centering
    \includegraphics[width=0.5\textwidth]{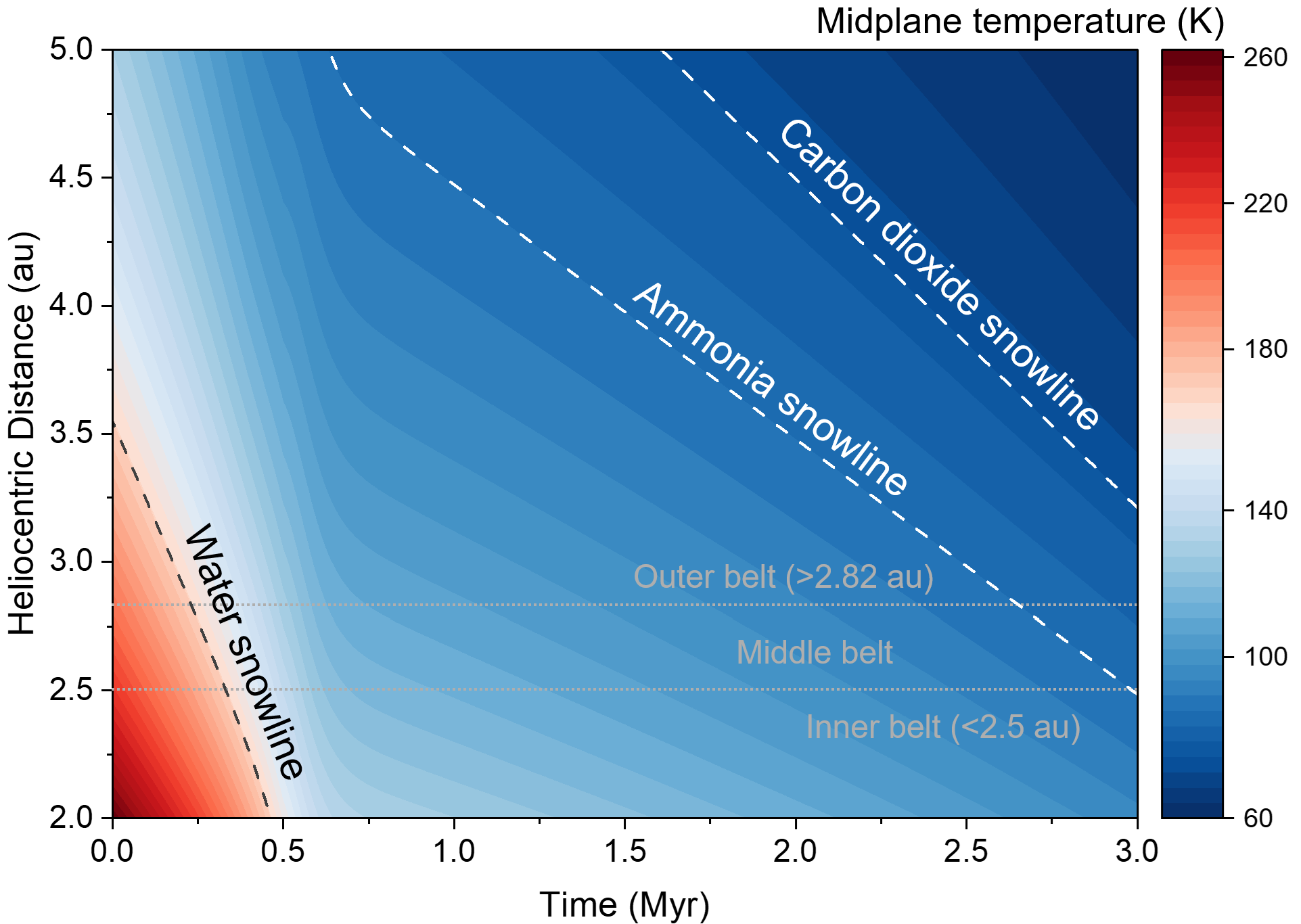} 
    \caption{Model of disk temperature evolution and snowline migration \citep{oka2011evolution} adapted in this study.}
    \label{fig:snowline}
\end{figure}

As noted before, this model does not directly simulate the evolution of the disk, and the growth and fragmentation of dust particles, but we still need a basic profile of the gas temperature $T$ in order to describe snowline migration and calculate the scale height of the gas disk and the azimuthal velocity of gas. Motivated by  previous studies of the thermal structure of protoplanetary disks and icy pebble accretion \citep{oka2011evolution, sato2016}, we simply use a power-law temperature profile $T(r,t)$, as given by,
\begin{equation}
    T(r,t) = T_{1\,\mathrm{au}}(t) \left( \frac{r}{1\,\mathrm{au}} \right)^{-\beta(t)}\, \mathrm{K},
\label{eq2}
\end{equation}

where $T_{1\,\mathrm{au}}(t)$ is the midplane temperature at 1\,au and $\beta(t)$ is the radial slope. The normalization $T_{1\,\mathrm{au}}(t)$ is calibrated so that snowlines follow simple migration tracks consistent with previous thermal evolution models \citep[e.g.,][]{oka2011evolution,mori2021evolution}. We consider a disk-evolution timescale of 3 Myr. Based on the results of \cite{oka2011evolution}, we assume an early viscous heating dominated phase for the first 0.5 Myr and adopt $\beta = 3/4$ \citep{1980MNRAS.191...37L} during which the disk cools rapidly. After this period, the disk becomes dominated by stellar irradiation, and the temperature profile gradually flattens. Thus, we change the power-law smoothly to $\beta = 1/2$ \citep{1981PThPS..70...35H}. In practice, $\beta(t)$ is prescribed to change smoothly from $3/4$ to $1/2$ between $t = 0.5$ and $0.7$ Myr with a hyperbolic tangent transition, and remains $\beta=1/2$ thereafter. This temperature profile allows the water snowline to migrate through the main asteroid belt at an early stage \citep[from 3.6 au to 2.0 au][]{oka2011evolution,mori2021evolution}. Snowlines for volatiles with lower condensation temperatures, such as ammonia and carbon dioxide, are initially located beyond Jupiter’s orbit. Over time, the ammonia snowline reaches $2.5\,$au and that of the carbon dioxide snowline reaches $3.2\,$au. The snowline migration scenario is illustrated in Figure~\ref{fig:snowline}.

We note that the temperature structure of the disk can have different radial dependence in reality. Under the control of the disk gas accretion, the boundary between the viscous-heating dominated and irradiation dominated region will gradually migrate inward, much like a snowline\citep{ida2019A&A...624A..28I}. In some cases, additional radiative effects, such as heated disk flares and reprocessing of radiation within the disk, could further reduce the temperature gradient, making it even shallower \citep[$T \propto r^{-3/7}$,][]{chiang1997spectral}. However, \cite{sato2016} demonstrated that icy pebble accretion onto an embryo at 1 $\mathrm{au}$ is largely unaffected by the specifics of the temperature profile. Since we are primarily concerned with the outer main belt region, where viscous heating is weaker, the details of the assumed temperature profile do not significantly impact our conclusions.

The azimuthal velocity of the gas is given by:
\begin{equation}
    v_{\phi,\mathrm{gas}} = (1 - \eta) v_\mathrm{K},
\end{equation}
where $\eta$ is the dimensionless parameter representing the pressure gradient in the gas disk \citep{nakagawa1986settling}:
\begin{equation}
    \eta = -\frac{1}{2} \left(\frac{c_\mathrm{s}}{v_\mathrm{K}}\right)^2 \frac{\partial \ln p}{\partial \ln r},
    \label{eq3}
\end{equation}

where $v_{\mathrm K}=r\Omega_{\mathrm K}$ is the Keplerian velocity and $p=\rho\,c_{\mathrm s}^2$ is the gas pressure at a given radius. The disk's turbulent strength is parameterized by $\alpha$ \citep{shakura1973black}, and it is defined as,
\begin{equation}
    \nu = \alpha c_\mathrm{s} h_\mathrm{g},
\end{equation}
where $\nu$ is the turbulent viscosity coefficient. The turbulent strength determines the scale height of the dust layer and, consequently, affects the accretion efficiency of pebbles onto planetesimals as explained below.

\subsubsection{Dust profile}

We employ a simplified approach to calculate the mass accretion flux of radially drifting pebbles (dust) onto planetesimals, assuming a steady pebble radial flux $\dot{M}_\mathrm{d}$ within the disk lifetime, with a fixed Stokes number $\mathrm{St}$. Previous studies show that the Stokes number remains nearly constant when dust particles drift while their size is limited by radial drift \citep{ida2016,taki2021new}. Pebbles in the fragmentation-limited state behave similarly, as the fragmentation limits the drifting pebble size, justifying our assumption of a fixed Stokes number in disk evolution. We discuss the limitations of this simplification in Section~\ref{3.2}.

The dimensionless stopping time is referred to as the Stokes number $\mathrm{St}$ describing the coupling between dust particles and the gas. The Stokes number is related to the stopping time $t_\mathrm{s}$ as,
\begin{equation}
    \mathrm{St} = \Omega_\mathrm{K} t_\mathrm{s}.
\end{equation}
We note that here the Stokes number can be written as \citep{sato2016},
\begin{equation}
    \label{eqst}
    \mathrm{St} = \frac{\pi}{2} \frac{\rho_\mathrm{p} a}{\Sigma_\mathrm{g}} \cdot \max\left(1, \frac{4a}{9\lambda_{\text{mfp}}}\right),
\end{equation}
where $\rho_\mathrm{p}$ is the material density of the pebbles, $a$ is the particle radius, and $\lambda_{\mathrm{mfp}}$ is the mean free path of gas molecules. Equation \ref{eqst} accounts for the different regimes of coupling, including Epstein and Stokes drag.

The radial mass flux of drifting pebbles $\dot{M}_\mathrm{d}$ is defined as,
\begin{equation}
    \dot{M}_\mathrm{d} = 2 \pi r  v_r  \Sigma_\mathrm{d}.
    \label{equst}
\end{equation}
The radial drift velocity of dust particles $v_r$ depends on the Stokes number, and is given by \citep{adachi1976gas},
\begin{equation}
    v_r = -\frac{2 \mathrm{St}}{1+\mathrm{St}^2} \eta v_\mathrm{K}. \label{eq:vr}
\end{equation}
With a given pebble radial flux $\dot{M}_\mathrm{d}$, the dust surface density $\Sigma_d(r)$ is calculated by \(\Sigma_d(r) = {\dot{M}}/({2 \pi r |v_r|})\). The vertical distribution of dust is influenced by gas turbulence and vertical settling, which leads to a scale height $h_\mathrm{d}$ of the dust layer \citep{youdin2007particle,okuzumi2012rapid}:
\begin{equation}
    h_\mathrm{d} = h_\mathrm{g} \left(1 + \frac{\mathrm{St}}{\alpha} \frac{1+2\mathrm{St}}{1+\mathrm{St}}\right)^{-1/2}.
\end{equation}
An additional effect of turbulence on stochastic pebble motion is ignored in this work; its potential impact on pebble accretion is discussed in Section~\ref{3.3}. The dust spatial density at the disk midplane can be expressed by, 
\begin{equation}
     \quad \rho_\mathrm{d} = \frac{\Sigma_\mathrm{d}}{\sqrt{2\pi}h_\mathrm{d}}.
\end{equation}

\subsection{Accretion Model}
\label{pebble}

The accretion rate $\dot{M}$ of pebbles onto planetesimals depends on parameters such as the particle density, turbulence strength, planetesimals mass, and the Stokes number $\text{St}$. As a planetesimal grows in mass, it passes through a sequence of accretion regimes characterized by different dynamics (see Figure~\ref{fig:ap}). Following the detailed analysis in previous studies \citep{ormel2010effect, morbidelli2012dynamics, lambrechts2012rapid, lambrechts2014forming, visser2016growth, Ormel2017}, we distinguish three types of accretion regimes: ballistic accretion, headwind pebble accretion, and shearing pebble accretion.

\subsubsection{Ballistic Accretion} 

At very low masses, ballistic regime (sometimes called the “drive-by” or no-settling regime) conditions apply. The planetesimal’s gravitational reach is too small to significantly deflect pebbles, their trajectories remain only weakly perturbed from the gas streamlines, i.e., they follow unbound, nearly hyperbolic paths past the body. Aerodynamic drag keeps the pebbles tightly coupled to the gas and prevents them from settling toward the planetesimal, so pebble accretion is negligible. This occurs when the planetesimal mass $M$ is below a critical value $M_1$. The threshold mass $M_1$ is defined as \citep{Ormel2017},
\begin{equation}
    M_1 = \frac{v_{\mathrm{hw}}^3 \mathrm{St}}{8 G \Omega_\mathrm{K}}, \label{eq:M1}
\end{equation}
where $v_{\mathrm{hw}} = \eta v_\mathrm{K}$ is the headwind velocity of disk gas against the planetesimal, in our model, equations~(\ref{eq2}) and (\ref{eq3}) together yield a headwind velocity of $v_{\mathrm{hw}} = 33\,\mathrm{m\,s^{-1}}$.

Given that the main objects of this study are asteroids, the reach of the gravitational and aerodynamic interactions between pebbles and planetesimals is small compared to the pebble layer’s thickness ($b \ll h_\mathrm{d}$, where $b$ is the effective impact parameter defined below for different regime), we evaluate the accretion rate following the established analytic model from \cite{visser2016growth} for planetesimals:
\begin{equation}
    \dot{M} = \pi R_\mathrm{p}^2 v_\infty \rho_d f_{\mathrm{coll}},
    \label{accretion1}
\end{equation}
where the relative velocity \( v_{\infty} \) is determined by both headwind and Keplerian shear, the latter resulting from the radial gradient in orbital velocity across the planetesimal’s encounter region. \( v_{\infty} \) is expressed as:
\begin{equation}
    v_{\infty} = v_{\mathrm{hw}} + \frac{3}{2} \Omega_{\mathrm{K}} b.
    \label{eq:v_shearing}
\end{equation}
At this stage, the headwind term dominates the relative velocity $v_{\infty} = v_{\mathrm{hw}}$. Finally, $f_{\mathrm{coll}}$ is the collisional efficiency, it can be calculated as follows:
\begin{equation}
    f_{\mathrm{coll}} = f_{\mathrm{coll,0}} \exp\left[a \left(\frac{\mathrm{St}}{\mathrm{St_*}}\right)^b\right].
    \label{collisional_efficiency}
\end{equation}
Here, $f_{\mathrm{coll,0}}$ is the uncorrected collision efficiency, and $a$ and $b$ are fitting parameters that depend on the adopted flow regime, as shown in Table~\ref{tab:coll_params}. In Table~\ref{tab:coll_params}, $\Theta$ is the Safronov number, defined as $\Theta = \left({v_{\mathrm{esc}}}/{v_{\mathrm{hw}}}\right)^2$ \citep{safronov1972evolution}.

\subsubsection{Headwind Pebble Accretion}

Once a planetesimal exceeds the critical mass threshold ($M > M_1$), drag-assisted accretion becomes significant, marking the transition from the ballistic to the headwind regime (also referred to as the Bondi regime in the context of pebble accretion). In this regime, the planetesimal's gravity can overcome the gas drag enough to capture pebbles and the relative velocity $v_{\infty}$ between the pebbles and the planetesimal is dominated by the headwind velocity $v_{\mathrm{hw}}$. This regime typically applies to intermediate-mass planetesimals where the planet’s Hill sphere is still small, or for pebbles with short stopping times. Following \citet{visser2016growth}, the accretion rate in the 3D headwind regime continues to follow Equations~\eqref{accretion1} and ~\eqref{collisional_efficiency}, but with different parameter choices specific to the headwind regime (Table~\ref{tab:coll_params}). 

We adopt the Stokes-flow regime in both the ballistic and headwind limits. \citet{visser2016growth} provides collision parameters for the potential- and Stokes-flow regimes. Although the large Reynolds numbers expected \citep{visser2016growth} indicate that the flow is inviscid, this does not readily mean that the flow is laminar (namely, the potential flow); inviscid approximation must break down near the planetesimal surface, where viscosity enforces a no-slip boundary and separation of the boundary layer produces turbulence \citep[e.g.,][]{landau1987fluid}. Thus, there is no consensus in literature which approximation is more appropriate; \citet{johansen2015growth} assumed the Stokes flow, while \citet{visser2016growth} argued that the gas flow is better described by potential flow. Given the uncertainty in the flow pattern around a planetesimal, we decided to assume the Stokes-flow regime because it maximizes the effective accretion cross-section for pebbles. If volatile delivery remains inefficient even under this optimistic assumption, it would be still more difficult under different flow models.

\begin{table}[ht]
    \centering
    \caption{Collision parameters for Stokes flow in the ballistic and 3D headwind regimes \citep{visser2016growth}.}
    \label{tab:coll_params}
    \begin{tabular}{lcccc}
        \hline
        Regime & $f_{\mathrm{coll,0}}$ & $\mathrm{St}_*$ & $a$ & $b$ \\
        \hline
        3D headwind & $2 \Theta \mathrm{St}$ & $2 \Theta + 4 + 4 / \Theta$ & $-2.26$ & $0.61$ \\
        Ballistic   & $1 + \Theta$           & $1 + \Theta$              & $3.24$  & $-0.86$ \\
        \hline
    \end{tabular}
\end{table}

When the planetesimal’s gravitational influence extends beyond the vertical thickness of the pebble layer (i.e., when the impact parameter $b$ exceeds the dust scale height $h_\mathrm{d}$), the system transitions to 2D accretion. In this case, pebbles approaching the planetesimal from all directions within the disk's full thickness can be captured. The 2D headwind accretion rate is given by,
\begin{equation}
    \dot{M}_{\mathrm{2D,hw}} = 2 b \Sigma_\mathrm{d} v_{\mathrm{hw}}.
    \label{eq:2D_headwind}
\end{equation}
The impact parameter \( b \) in the headwind regime follows:
\begin{equation}
    b_{\mathrm{hw}} = \sqrt{\frac{2 G M \mathrm{St}}{v_{\mathrm{hw}}^2 \Omega_\mathrm{K}}},
    \label{eq:b_headwind}
\end{equation}
which determines the effective cross-section for pebble accretion.

\subsubsection{Shearing Pebble Accretion}

For larger planetesimals, their encounter dynamics shift to the shearing regime, where Keplerian shear becomes the dominant force governing relative motion. This shear-dominated accretion corresponds to what is sometimes called the “Hill regime” in the pebble accretion theory \citep{ormel2010effect}. The transition occurs when the planetesimal mass surpasses a second threshold \( M_2 \), defined as,
\begin{equation}
    M_2 = \frac{v_{\mathrm{hw}}^3}{8 G \Omega_\mathrm{K} \mathrm{St}}.
\end{equation}
At this stage, the relative velocity \( v_{\infty} \) is determined by Keplerian shear rather than the headwind $v_{\infty} = v_{\mathrm{hw}} + \frac{3}{2} \Omega_{\mathrm{K}} b$.

The pebble accretion rate in the 3D shear regime is given by,
\begin{equation}
    \dot{M}_{\mathrm{3D}} = \pi b^2 \rho_d v_{\infty},
\end{equation}
where \( b \) is the impact parameter, now determined by,
\begin{equation}
    b_{\mathrm{sh}} = \left(\frac{GM_\mathrm{p} t_{\mathrm{stop}}}{\Omega_\mathrm{K}}\right)^{1/3}
\end{equation}

In the 2D shearing regime, the planetesimal’s capture radius spans the entire disk height, leading to more efficient accretion. The accretion rate follows:
\begin{equation}
    \dot{M}_{\mathrm{2D,sh}} = 2 b \Sigma_\mathrm{d} v_{\infty}.
    \label{eq:2D_shearing}
\end{equation}
This expanded capture radius effectively increases the probability of capturing pebbles drifting through the disk, enhancing the planetesimal’s growth rate.

\subsubsection{Accretion Efficiency} 

To quantify what fraction of the radial pebble flux is accreted by the planetesimal, we define the pebble accretion efficiency \(\epsilon_{\mathrm{acc}}\) as \citep[also called accretion probability][]{Ormel2017}:
\begin{equation}
    \epsilon_{\mathrm{acc}} = \frac{\dot{M}}{\dot{M}_{\mathrm{d}}} = \frac{\dot{M}}{2 \pi r \, v_{r} \, \Sigma_\mathrm{d}},
\end{equation}
where \(r\) is the orbital radius, \(v_{\mathrm{drift}}\) is the radial drift speed of pebbles, and \(\Sigma_p\) is the pebble surface density. Note that a pebble accretion efficiency exceeding unity is unrealistic. Thus, we enforce,
\begin{equation}
    \epsilon_{\mathrm{acc}} \leq 1
\end{equation}
Under this condition, the maximum accretion rate is given by: $\dot{M}_\mathrm{max} = 2 \pi r \, \Sigma_\mathrm{d} \, v_r$.

\begin{figure}[htbp]
    \centering
    \includegraphics[width=0.8\textwidth]{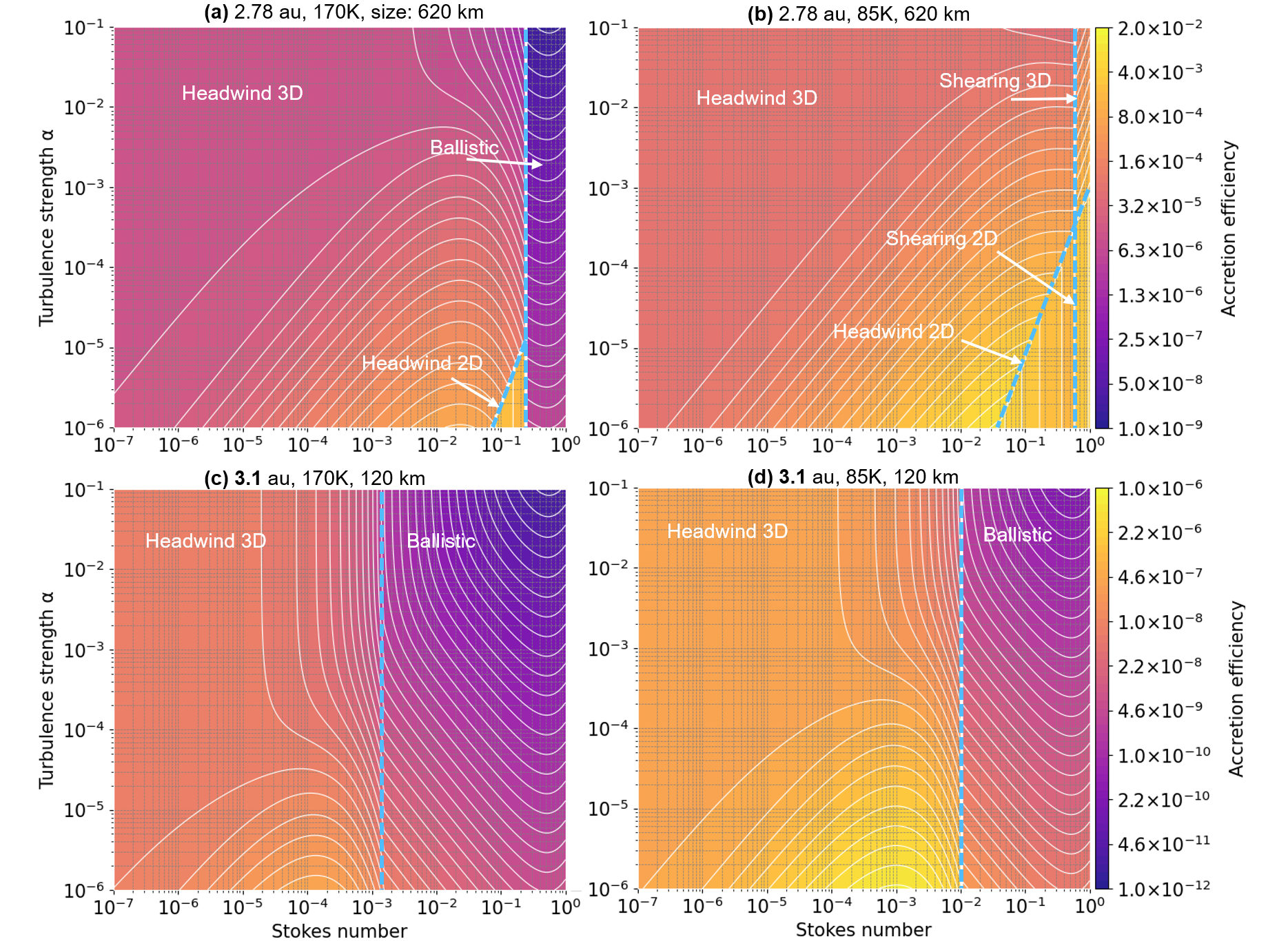}
    \caption{
    Pebble accretion efficiency $\epsilon_{\rm acc}$ (color map with white contours) as a function of turbulence strength $\alpha$ and Stokes number ${\rm St}$ for two representative-sized planetesimals at two midplane temperatures.  
    Panels (a)–(b) show a 620\,km planetesimal at 2.78\,au (Ceres embryo) for $T_{\rm mid}=170$\,K and 85\,K, respectively;  
    panels (c)–(d) show a 120\,km planetesimal at 3.1\,au under the same two temperatures.  
    The light-blue dashed curves mark the boundaries among the accretion regimes. Notably, smaller planetesimals require significantly smaller Stokes numbers ($\mathrm{St} \lesssim 10^{-3}$) to enter efficient accretion regimes, and lower temperatures could shift the boundaries and enable higher accretion efficiencies.
    }
    \label{fig:ap}
\end{figure}

Figure~\ref{fig:ap} illustrates how the pebble accretion efficiency $\epsilon_{\mathrm{acc}}$ varies in the \((\alpha,\,\mathrm{St})\) parameter space for planetary embryos of different sizes at various orbital locations. We observe that smaller embryos are predominantly governed by 3D accretion; however, at large Stokes numbers, they enter the ballistic (collision-dominated) regime, resulting in lower accretion probabilities. In contrast, larger embryos can efficiently accrete pebbles in all regimes. In particular, very large embryos (e.g., those in the final stages of Jupiter's core growth) transition to the 2D shear-dominated regime, where the accretion efficiency $\epsilon_{\mathrm{acc}}$ approaches unity. For the intermediate-sized MBAs considered in this study ($D \simeq 120$--$200\ \mathrm{km}$), their accretion process mainly falls within the ballistic regime (dominated by collisions) and the 3D headwind regime, depending on the specific Stokes number and size of the planetesimals. We note, however, that for larger bodies (e.g., terrestrial embryos or giant-planet cores), readers should consult the revised expressions of pebble accretion efficiency with updated coefficients compiled in \cite{OL18,ormel2024}.

\subsection{Assumptions and Constraints}
\label{assumptions}

\subsubsection{Input Parameters and Their Ranges}

We set the pebble radial flux $\dot{M}_{\mathrm{d}}$, the Stokes number $\mathrm{St}$, and the turbulence strength $\alpha$ as free parameters for the calculation of pebble accretion. Specifically, we consider $\dot{M}_{\mathrm{d}}$ from $10^{-1} \, \mathrm{M_\oplus / Myr}$ to $10^{3} \, \mathrm{M_\oplus / Myr}$, $\alpha$ from $10^{-6}$ to $10^{-1}$, and $\text{St}$ from $10^{-7}$ to $1$ \footnote{Although the terms \textquotedblleft pebble\textquotedblright\ and \textquotedblleft pebble accretion\textquotedblright\ are generally used for particles with $10^{-3}\lesssim\mathrm{St}\lesssim10^0$ in literature \citep[e.g.,][]{Ormel2017}, we use those terms for particles with a wider range of $\mathrm{St}$ for the sake of simplicity.}, chosen to encompass the full range of possible disk conditions and pebble accretion scenarios explored in previous theoretical studies \citep[e.g.,][]{johansen2017forming, armitage2020astrophysics, johansen2021pebble}.

We note that fixing the Stokes number means that the pebble size varies with the heliocentric distance. Given that the Stokes number is defined as $\mathrm{St} \propto a \Sigma_{\mathrm{g}}^{-1}$ (Equation (\ref{eqst})) and the gas surface density follows $\Sigma_{\mathrm{g}} \propto r^{-3/2}$, the physical radius of pebbles in our model follows $s \propto r^{-3/2}$. This implies that, at larger distances from the Sun, pebbles are physically smaller even if their aerodynamic properties remain unchanged. Such a size–distance trend is expected for drifting pebbles in the Epstein regime, where they tend to maintain a roughly constant Stokes number during their inward migration \citep[see e.g.,][]{ida2016}. Therefore, assuming a fixed Stokes number is consistent with the expected physical evolution of pebbles in protoplanetary disks. Additionally, we assume that the planetesimals are already formed and do not undergo significant migration during the accretion process.

\subsubsection{Icy Pebble and Snowline Migration}

With reference to the cometary composition \citep{mumma2011chemical}, we consider icy pebbles with a density of $\rho_\mathrm{p}=1.4\,\text{g\,cm}^{-3}$ and a rock-to-ice mass ratio of 1:1. We assume the ice fraction consists primarily of $79\%$ water, $1\%$ ammonia, and $20\%$ carbon dioxide by mass, following observed cometary compositions \citep{mumma2011chemical}. Given that water is the dominant volatile component, we assume that all volatile materials sublimate once the pebble crosses the water snowline, leading to a 50\% mass loss. For simplicity, we neglect any additional mass change that may occur at the ammonia and carbon dioxide snowlines. Between the ammonia and carbon dioxide snowlines, the volatile composition consists of 98.9\% water and 1.1\% ammonia, and 100\% water between the water and ammonia snowlines. Since the mass \(m\) of a spherical particle scales as \(m \propto a^3\), halving the mass reduces the pebble radius \(a\) by a factor of \((0.5)^{1/3}\) and the new particle radius becomes $a_{\mathrm{new}} \;=\; a_{\mathrm{old}}\,(0.5)^{1/3}$. Consequently, as the pebble crosses the snowline, it becomes more tightly coupled to the gas, and its Stokes number decreases. Based on the definition of the Stokes number in different drag regimes (Equation (\ref{eqst})), in the Epstein regime (\( a < 9\lambda_{\mathrm{mfp}} / 4 \)), the Stokes number scales as,
\begin{equation}
    \mathrm{St}_{\mathrm{new}} = (0.5)^{1/3} \, \mathrm{St}_{\mathrm{old}}.
\end{equation}
In the Stokes regime, where the particle size exceeds the mean free path (\( a > 9\lambda_{\mathrm{mfp}} / 4 \)), the Stokes number follows a different scaling:
\begin{equation}
    \mathrm{St}_{\mathrm{new}} = (0.5)^{2/3} \, \mathrm{St}_{\mathrm{old}}.
\end{equation}
When pebbles lose mass due to sublimation, their Stokes number decreases significantly. This decrease enhances the coupling between pebbles and the gas, resulting in a reduction of the radial drift velocity and, consequently, a longer residence time in the disk before accretion onto planetesimals. 

Also, we note that this is a simplification. In reality, silicate pebbles are denser and smaller in grain size. Sublimation of the icy mantle of a grain or a single ice–rock aggregate may release multiple small rocky grains with significantly different aerodynamic properties. Several studies have investigated such processes near the snowline, focusing on the release, fragmentation, and redistribution of silicate grains, as well as the formation of planetesimals inside the water snowline, with more realistic treatments, \citep[e.g.,][]{Ros&Johansen2013,ida2016,Schoonenberg2017,hyodo2021,wang2025,sirono2025}. We acknowledge that, under specific conditions, snowlines-related dust pile-up mechanisms may provide additional impetus for ice transport on asteroids. We discuss the potential influence of dust pile-up near the snowlines and its implications for pebble delivery on asteroids in Section \ref{3.1}.

\subsubsection{Initial Planetesimal Seeds} \label{subsubsec:seeds}

At \( t = 0 \,\text{Myr} \), we initialize a population of rocky planetesimals with sizes ranging from 100\,km to 1000\,km in diameter. These planetesimals act as the seeds for further growth via the pebble accretion. The accretion process is assumed to last for \( 3 \,\text{Myr} \), consistent with the typical lifetime of the protoplanetary disk \citep{armitage2020astrophysics} and the estimated accretion timescales of chondritic parent bodies \citep{connelly2012absolute}. Among these planetesimals, those in the asteroid belt region (2--5\,au) are modeled with diameters ranging from 120\,km to 620\,km. The size (diameter) range was selected because this study focuses on large main-belt asteroids. An assessment of the size distribution of main-belt asteroids indicates that asteroids with a diameter exceeding 120 km are likely to be primitive and not remnants of a parent body that has undergone a catastrophic impact \citep{bottke2005linking}. We consider the largest planetesimal in our model to be the embryo of the largest asteroid in the main belt, Ceres, whose current water content reaching approximately 30\% \citep{prettyman2017extensive}. As we assume that Ceres acquired its water from icy pebble accretion, its rocky embryo should have 40\% of its present-day mass. 

We assume the rocky planetesimals have an average bulk density of \( 3\,\text{g\, cm}^{-3} \) as the density of typical large S-complex asteroids \citep{carry2012density, vernazza2021vlt}, whereas real C-complex asteroids are likely more porous, with an average bulk density of $\simeq2\,\mathrm{g\,cm^{-3}}$. This implies that their actual gravitational ability to accrete pebbles is lower than our assumed case, effectively relaxing the constraints on the required Stokes number for efficient accretion.

\subsubsection{Constraints to Test the Icy Pebble Accretion Model}
\label{Constraints}

We adopt two primary constraints to evaluate whether pebble accretion can successfully grow planetesimals and embryos to conform to the current Solar System: compositional and mass constraints.

We define the compositional (morphological) constraint as a set of minimum thickness estimates for the water-bearing layer (containing only water ice), ammonia-bearing layer (including both water and ammonia ices), and a carbon dioxide-bearing layer (composed of water, ammonia, and carbon dioxide ices) that asteroids must have accreted to be compatible with current observations. In practice, this constraint is informed by surface spectral data and estimates of impact excavation depths, linking the observed spectral homogeneity to the distribution of volatiles beneath the surface. Currently, limited observations of large main-belt asteroids provide key insights into their compositional trends with heliocentric distance \citep{demeo2015compositional, rivkin2022nature}. Although reflectance spectra reveal only the composition of the asteroid’s surface, leaving the bulk composition largely unknown, but in situ and astronomical observations of main-belt asteroids have detected the presence of a 3 $\mu$m band, suggesting that volatile-bearing materials are widespread across their surfaces. Data from the Dawn mission reveal a global 3.1 $\mu$m absorption feature on Ceres, indicating that the stratification of the ammonia layer extends at least several tens of kilometers \citep{ammannito2016distribution}. For other large asteroids ($D > 120 \ \mathrm{km}$) which lack in situ measurements, multiple observations at different phases show stable spectral features of 2.7 $\mu$m and/or 3.1 $\mu$m \citep{ammannito2016distribution,usui2019akari,rivkin2022nature, takir2023late}, with no significant spectral type transitions observed \citep{rivkin2022nature}. Given the spectral homogeneity observed across large asteroids, we assume that volatiles and hydrous minerals extend at least to typical depths of impact crater excavation. 

Based on observational data cross the Solar System small bodies and satellites, \citet{leliwa2008impact} provided an estimate for the maximum impact crater diameters $D_\mathrm{max}$ on both the icy and rocky objects:
\begin{equation}
    D_\mathrm{max} = 0.45 \cdot D  \quad \text{for } D < 420 \text{ km},
\end{equation}

Then, the excavation depth of a crater depends on the cratering flow field from the transient crater, the short-lived, initial cavity formed immediately after the impact before collapse or modification.  Generally, the transient crater excavation depth $d_\mathrm{e}$ is approximately 10\% of the crater diameter $D_\mathrm{t}$:
\begin{equation}
    d_\mathrm{e} / D_\mathrm{t} \approx 0.1.
\end{equation}
The above relationship is valid for both simple craters, which retain a bowl-shaped morphology, and complex craters---larger craters that undergo structural collapse and exhibit features like central peaks, stepped walls, and collapse rings---and is a property common to most craters on different bodies \citep{croft1980cratering,melosh1989impact}. A transient crater undergoes collapse and eventually forms the final crater. We adopt the following empirical relationships between transient crater diameter ($D_\mathrm{t}$) and final crater diameter ($D_\mathrm{f}$):
for simple craters \citep{shoemaker1983asteroid},
\begin{equation}
    D_\mathrm{t}  = 1.25 \cdot D_\mathrm{f},
\end{equation}
and for complex craters \citep{croft1985scaling},
\begin{equation}
    D_\mathrm{t} = D_\mathrm{Q}^{0.15} \cdot D_\mathrm{f}^{0.85},
\end{equation}
where $D_\mathrm{Q}$ represents the simple-complex transition diameter, which depends on the body's surface gravity and material properties \citep{schenk2004ages}. For Ceres, this transition occurs at approximately $D_\mathrm{Q} \sim 10$ km \citep{hiesinger2016cratering}. Given that most C-complex MBAs with 100 to 200 km diameters have lower surface gravity than Ceres, the transition diameter may be slightly larger (see Figure~2b of \cite{hiesinger2016cratering}). As this increase is expected to be minor, so we adopt $D_\mathrm{Q} = 10$ km for all asteroids, ensuring that the largest craters on all asteroids are complex craters. Thus, for a given asteroid diameter $D$, the maximum excavation depth exposing the deepest subsurface materials is given by,
\begin{equation}
    H = 0.1 \cdot D_\mathrm{Q}^{0.15} \cdot (0.45 \cdot D)^{0.85},
\end{equation}
which can be approximated as,

\begin{equation}
    \left(\frac{H}{1\ \mathrm{km}}\right)\;
    \approx\;0.07\,
    \left(\frac{D}{1\ \mathrm{km}}\right)^{0.85}.
\end{equation}

To ensure that the icy layer accumulated through icy pebble accretion fully covers the asteroid’s surface, we impose this excavation depth $H$ as a constraint, requiring that the total icy layer thickness, $\Delta H_\mathrm{ice}$, satisfies: $\Delta H_\mathrm{ice} > H$. Specifically, $\Delta H_\mathrm{ice}$ includes contributions from both water and ammonia ices, while $\Delta H_\mathrm{NH_3}$ represents the thickness of the ammonia-bearing icy layer. It is important to note that this constraint only sets a lower bound, and the actual volatile content may be significantly higher because of the necessary conditions for onset of aqueous alteration and post-alteration processes. Studies of hydrated CCs and aqueous alteration suggest that their parent bodies probably experienced high bulk water-to-rock ratios (W/R) and internal mixing \citep[e.g. ][]{suttle2021aqueous, kurokawa2022distant} to allow extensive alteration. Furthermore, as various post-hydration processes—including space weathering, degassing, and impact heating could have led to substantial volatile loss over time \citep{suttle2021aqueous}. This implies that the actual volatile mass accreted by asteroids needs to be much higher than the threshold set by the topographic constraint. Nevertheless, we adapt the above-mentioned compositional constraint as our methodology is designed to test whether icy pebble accretion can satisfy the most relaxed conditions for MBAs.

For the mass constraint, we assume that planetesimals (embryos) should grow to within approximately $\pm 10\%$ of their expected final masses. We assume the largest main belt body in our model, with its initial rocky embryo accounting for 40\% of its present-day mass, should grow to a size comparable to Ceres with $\sim 30~\mathrm{wt}\%$ of water. These
mass thresholds represent the required outcomes of pebble accretion under optimal conditions and serve as benchmarks for assessing the efficiency of planetesimal growth in the asteroid belt region.

Together, these two constraints provide a framework for evaluating the feasibility of pebble accretion as a mechanism for planetesimal and embryo growth: (1) The compositional constraint ensures that an accreted icy layer (water, ammonia, and carbon dioxide) is sufficiently thick to cover surface topographic variations, thereby preserving the volatile-rich composition inferred from observations. (2) The mass constraint ensures that growing bodies reach the expected sizes required for their respective roles in planetary formation. By applying both criteria, we can systematically assess whether pebble accretion can account for the observed diversity of bodies in the Solar System.

\subsection{Summary of Workflow}

We investigated the growth of planetesimals and embryos via icy pebble accretion, assuming a steady radial flux of pebbles in a static protoplanetary disk with migrating snowlines. To evaluate the viability of pebble accretion under various conditions, we applied (1) compositional constraint and (2) mass constraint as two key constraints to evaluate the results.

We then assessed whether such conditions could have been achieved in the primordial protosolar disk. Results for asteroid accretion are presented in Section~\ref{sec:results}, while the conditions for gas giant core formation and the possible scenarios of Solar System evolution with pebble accretion are discussed in Section~\ref{sec:discussion}.

\section{Results} \label{sec:results}

Here we investigate the potential of icy pebble accretion for volatile accretion on MBAs. To explore the three-dimensional parameter space defined by the Stokes number $\mathrm{St}$, turbulence strength $\alpha$, and the pebble radial flux $\dot{M}_\mathrm{d}$, we first derive $\dot{M}_\mathrm{d}$ which satisfies the conditions for sufficient growth of Ceres (the largest and primordial MBA) from a rocky embryo as a function of $\mathrm{St}$ and $\alpha$. Then, we show the growth of planetesimals in the main belt under the derived pebble radial flow, $\dot{M}_\mathrm{d}(\mathrm{St},\alpha)$. Finally, we evaluate the accumulation of icy layers $\Delta H$ on the surfaces of asteroids by applying the derived pebble radial flux $\dot{M}_\mathrm{d}(\mathrm{St},\alpha)$. 

\begin{figure}[htbp] 
    \centering
    \includegraphics[width=1\textwidth]{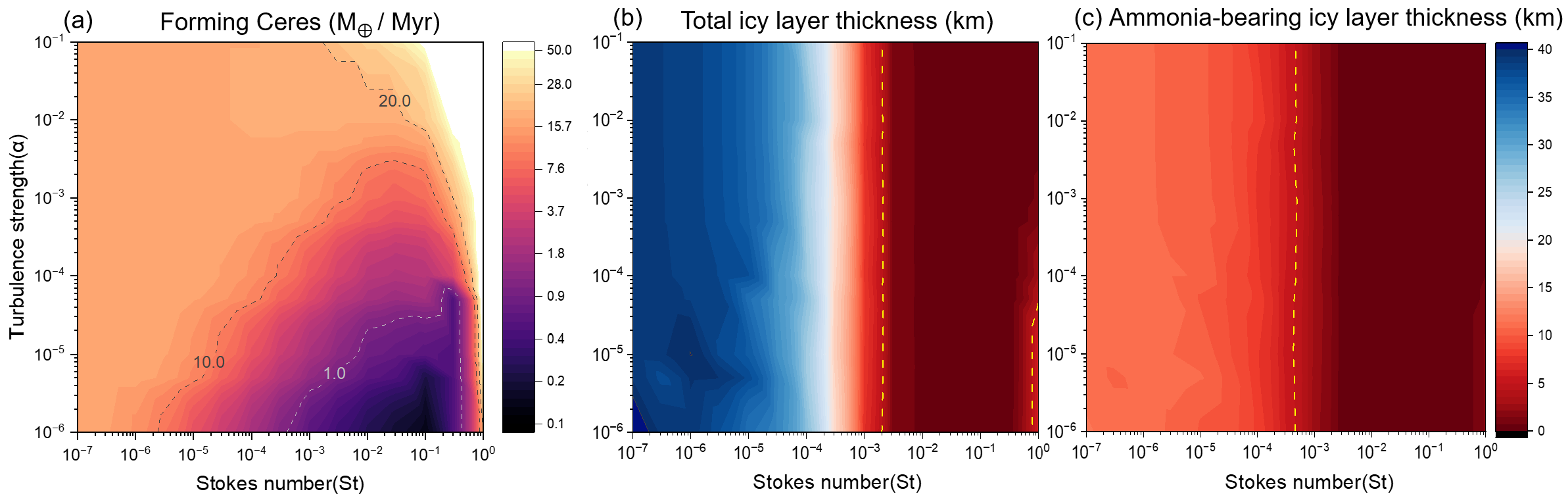} 
    \caption{(a) The required pebble radial flux $\dot{M}_\mathrm{d}$ for forming Ceres as a function of $\mathrm{St}$ and $\alpha$. (b) Total icy layer thickness $\Delta H_\mathrm{ice}$ on a 120 km sized asteroid at 2.7 au, including both H$_2$O- and NH$_3$-bearing ices, accreted at different orbits derived from $\dot{M}_\mathrm{d}(\mathrm{St}, \alpha)$. (c) Thickness of the NH$_3$-bearing icy layer $\Delta H_\mathrm{NH_3}$ of a same-sized asteroid at 3.1 au. The yellow dashed lines in (b)-(c)  mark the thickness constraint $\Delta H = 4.1\,$km.}
    \label{result1} 
\end{figure}

 Figure~\ref{result1}a illustrates the pebble radial flux $\dot{M}_\mathrm{d}$ required to form Ceres. In contrast to large pebble radial flux ($\dot{M}_\mathrm{d}\sim 10^2\ M_\oplus/\mathrm{Myr}$) required for the growth of gas giant cores and other Solar System planets \citep[e.g.,][]{johansen2017forming,morbidelli2015great,johansen2021pebble,chambers2023making}, the growth of Ceres from its rocky embryo requires a moderate pebble radial flux of $\dot{M}_\mathrm{d}\sim 10~M_\oplus/\text{Myr}$ and it is feasible to supply sufficient amounts of water and ammonia ice to Ceres via icy pebble accretion under realistic disk conditions. Under most conditions, the accretion process primarily follows the 3D headwind accretion regime (see Figure~\ref{fig:ap}a). At higher turbulence strengths ($\alpha \gtrsim 10^{-2}$), the accretion efficiency of pebbles onto Ceres embryo is low due to the increased dust scale height, which reduces the dust spatial density at the midplane. Additionally, the critical mass $M_1$ for the transition from the ballistic to the headwind regime increases as $\mathrm{St}$ increases (Equation (\ref{eq:M1})); this further reduces the accretion efficiency onto the embryo, requiring a higher pebble flux to form Ceres—up to $18~M_\oplus/\text{Myr}$ in the case of $\mathrm{St} > 10^{-2}$. For $\alpha \leq 10^{-3}$, a larger Stokes number results in significantly higher accretion probabilities for embryos, as pebbles with weaker gas coupling experience reduced aerodynamic drag and maintain higher relative velocities, increasing the accretion cross section
\citep{lambrechts2014forming,johansen2015growth,Ormel2017,OL18}. For $\alpha \leq 10^{-4}$ and $\mathrm{St} \geq 10^{-3}$, the required pebble flux is less than $1\,M_\oplus\,\mathrm{Myr}^{-1}$. As we explain below, with the constrained $\dot{M}_\mathrm{d}(\mathrm{St},\alpha)$, Ceres acquires a sufficient amount of ammonia ice ($\Delta H \sim 25~\mathrm{km} $) to be consistent with our compositional constraints. Variations in Stokes number and $\dot{M}_\mathrm{d}$ primarily affect the thickness of ammonia-bearing ice layers on asteroids smaller than $\simeq 300$ km, while the impact on largest asteroids is marginal.

Under the conditions suitable for Ceres formation ($\dot{M}_\mathrm{d}(\mathrm{St},\alpha)$), Figure~\ref{result1}b illustrates the thickness of icy layers formed on an asteroid with an initial size of 120 km located at 2.7~\text{au}, a typical position for hydrated asteroids in the middle main belt \citep{rivkin2022nature, takir2023late}. The results show that the thickness of the water-bearing icy layer is relatively insensitive to $\alpha$ but requires $\mathrm{St}\lesssim10^{-3}$ to allow sufficient volatile accretion to meet the topographic constraints ($\Delta H\geq 4.1\ \mathrm{ km}$). 

Supplying sufficient amounts of ammonia ice to the outer MBAs is more difficult. Figure~\ref{result1}c shows the thickness of ammonia-bearing icy layers for a 120~km asteroid located at 3.1~au, a typical position for ammonia-bearing asteroids in the outer main belt \citep{rivkin2022nature, takir2023late}. The ammonia-bearing icy layer is significantly thinner than that of water ice, requiring $\mathrm{St}\ \lesssim 10^{-4}$) to satisfy same constraint. For $\text{St} \gtrsim 10^{-3}$, only a thin ammonia-containing veneer forms on the asteroid's surface, with a thickness of just a few tens of meters. This is because the accretion efficiency of pebbles decreases with increasing heliocentric distance, and the ammonia snowline reaches the main belt much later than that of water.

\begin{figure}[htbp] 
    \centering
    \includegraphics[width=1\textwidth]{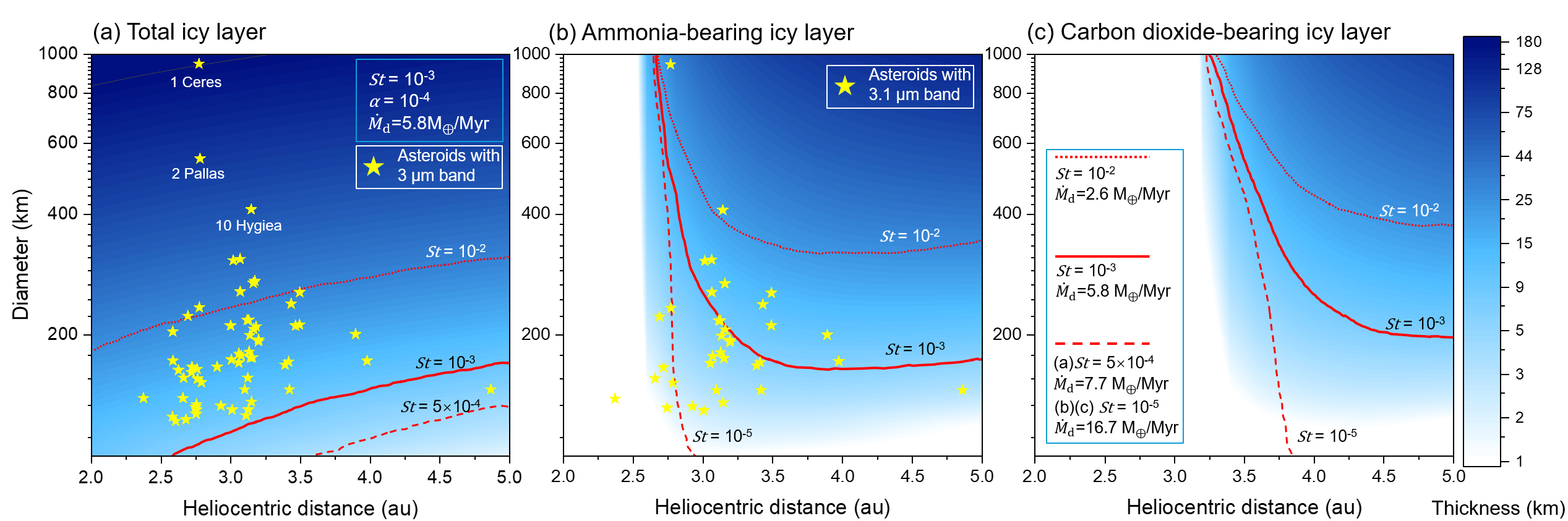} 
    \caption{ (a) Final total icy layer $\Delta H_\mathrm{H_2O}$, (b) ammonia-bearing icy layer $\Delta H_\mathrm{NH_3}$, (c) carbon dioxide-bearing icy layer $\Delta H_\mathrm{CO_2}$ thickness as a function of heliocentric distance and asteroid size, compared with large MBAs exhibiting these features. Red solid lines show the boundaries above which $\Delta H$ meets compositional constraints under the same parameter, and the dashed and dotted lines represent the boundaries for results using different St and $\dot{M}_\mathrm{d}$.} 
    \label{result2} 
\end{figure}

Detailed comparison to existing MBAs exhibiting $3\ \mathrm{\mu m}$ bands confirms that small St of pebbles are required to supply ammonia via icy pebble accretion, especially to smaller bodies ($D\lesssim 200\ \mathrm{km}$). Figure~\ref{result2} shows the thickness of volatile-rich layers obtained for a range of heliocentric distances and asteroid sizes and its comparison to existing MBAs. As discussed above, with $\dot{M}_\mathrm{d}(\mathrm{St},\alpha)$ that allows for Ceres' formation, the thickness of the icy layer $\Delta H$ on a $120\ \mathrm{km}$ asteroid is predominantly controlled by $\mathrm{St}$, while $\alpha$ has a negligible effect. Based on this, we select a typical turbulence strength of $\alpha=10^{-4}$ and choose $\mathrm{St}$ values which meet the constraints for water and ammonia thickness on 120~km asteroids. This allows us to systematically investigate pebble accretion across the asteroid belt and compare the results with observations. The results shown in Figure~\ref{result2} indicate a strong dependence of icy layer thickness on planetesimal size, with larger planetesimals developing substantially thicker volatile-rich layers. The relationship between the thickness of the icy layer and heliocentric distance differs for water and ammonia. Although the thickness of the water ice layer (Figure~\ref{result2}a) exhibits a weak dependence on the heliocentric distance, the thickness of the ammonia (Figures~\ref{result2}b) and carbon dioxide(Figures~\ref{result2}c) ice layer increases significantly with distance due to the longer time span available for the icy pebbles to accrete in the outer region. A comparison with observations of large C-complex asteroids reveals that, even with $\text{St} = 10^{-5}$, only approximately 80\% of asteroids could accrete enough ammonia to match the constraint (solid line on panel (b)), despite the ammonia snowline having actually migrated to within 2.5 au. However, for $\text{St} = 10^{-3}$, approximately 30\% asteroids are capable of accreting a sufficiently thick ammonia-bearing icy layer (panel (b) in Figure~\ref{result2}), although nearly all asteroids could accrete water ice layer meeting the required topographic constraint.

The accretion of carbon dioxide ice is more challenging. Under our Ceres–calibrated pebble flux, no asteroid in the main belt region is able to build carbon dioxide–bearing icy layer thick enough to satisfy the compositional constraint (Figure~\ref{result2}c). This is consistent with the absence of carbon dioxide features among MBAs, although non-detections do not uniquely imply the absence of carbon dioxide accretion, because subsequent volatile loss and surface processing could also erase these features.

\section{Discussion} \label{sec:discussion}

\subsection{Constraints on the Required Small Dust Sizes}
\label{D1}
 
Although our model is optimized for icy pebble accretion onto MBAs (Section \ref{subsubsec:seeds}), which includes a relaxed assumption of high planetesimal densities, it nonetheless implies that relatively small Stokes numbers are needed to build up the inferred ice layers on $100$--$200$~km bodies (Section~\ref{sec:results}): $\mathrm{St} \lesssim 10^{-3}$ for water and $\mathrm{St} \lesssim 10^{-4}$ for ammonia. At $r \simeq 3.1$ au, these values correspond to grain sizes ranging from a hundred micrometers to a millimeters. In this subsection we discuss to what extent such small pebbles are compatible with, or in tension with, theoretical and observational constraints on dust evolution in protoplanetary disks, and what this implies for the in-situ pebble-accretion scenario. Here we introduce constraints from i) dust-to-gas ratio, ii) theoretical prediction for dust coagulation and iii) protoplanetary disk observations.

First, the slower drift requires unrealistically high dust-to-gas ratios ($Z \gtrsim 1$, compared to the typical disk value of $\sim 0.01$), estimated as $Z = \Sigma_{\mathrm{d}}/\Sigma_{\mathrm{g}}$, where $\Sigma_{\mathrm{d}}$ is inferred from the assumed pebble flux $\dot{M}_\mathrm{d}$ and radial drift velocity $v_\mathrm{r}$ (see Equations~\ref{eqst}~and~\ref{eq:vr} in Methods). To sustain the pebble flux assumed in our model, maintaining a sufficiently high dust-to-gas ratio (for $\text{St} \sim 10^{-5}$, $Z \gtrsim 1$) is necessary. However, in such high-dust environments, self-gravity can trigger streaming instability or even direct gravitational collapse into planetesimals, rather than supporting prolonged pebble accretion \citep{toomre1964gravitational,goldreich1973formation,johansen2007rapid}.

The second case is from theoretical studies on dust coagulation in protoplanetary disks. Given that dust grains in protoplanetary disks originate from the interstellar medium, where grain sizes predominantly range from $0.1$ to $1\,\mu$m \citep{draine2003interstellar}, such small particles are expected to rapidly grow into larger aggregates until reaching fragmentation thresholds or encountering bouncing barriers \citep{armitage2020astrophysics}. Beyond the water snowline, where the sticking efficiency of dust particles may be higher, theoretical studies predict that pebbles grow to much larger Stokes numbers \citep[$\mathrm{St} > 10^{-2}$,][]{yap2024dust}, although some experiments suggest that water ice may not be sticky in low temperatures expected in protoplanetary disks \citep{Musiolik2019ApJ...873...58M}. In addition, in the outer main belt region through which the ammonia snowline passed, icy pebbles may be covered by mantles with ammonia, whose surface properties could significantly affect their stickiness and collisional outcomes. However, direct experimental data on its stickiness are scarce, but some studies suggest that carbon dioxide ice may be less sticky than water ice at low temperatures \citep{2016ApJ...818...16M}, potentially inhibiting efficient pebble growth in these regions, leading small dust size.

Finally, while observations of protoplanetary disks generally indicate that dust grains tend to grow to relatively large sizes, with sub-millimeter to centimeter-sized particles dominating the dust mass budget \citep{wilner2005toward,perez2012constraints,testi2014dust,andrews2020observations,li2023radial}, recent studies offer a more nuanced view. Analysis of scattering albedo of the disk suggest that in the inner regions the maximum grain size can be as small as $\sim 340\,\mu$m, corresponding to ${\rm St} \sim 10^{-4}$ \citep{ohashi2023dust,yoshida2024dust}. Such particles would align with our requirements for water and ammonia delivery to most of the outer main belt asteroids, although they would need to be maintained at relatively high dust surface densities or longer timescales for accretion \citep{OH2025}.

In summary, both theoretical models and observational constraints indicate that maintaining a sufficiently high population of very small pebbles is non-trivial. While particles with ${\rm St} \sim 10^{-4}$ may plausibly exist and could account for volatile delivery to $100$--$200$~km asteroids, the feasibility of achieving this condition within the protoplanetary disk remains to be explored. Thus, our results suggest that in situ icy pebble accretion can provide a possible pathway for volatile delivery to MBAs, but that building up the observed ammonia-bearing ice layers in this way remains challenging.  Further exploration beyond the simple scenarios discussed in this paper may be required to understand the icy pebble accretion on asteroids and the subsequent evolutionary processes of accreted icy layers (melting, alteration, dehydration, etc.)

\subsection{Implications for Future Observations and Studies of Volatile-Rich Asteroids}
\label{D3}

Our results indicate that the inward drift of icy pebbles provides only a limited mechanism for delivering volatiles to large MBAs. While most asteroids may have acquired water-bearing icy layer through this process, sufficient amounts of ammonia appear to be accreted only by the largest bodies ($\gtrsim 200$ km), such as Ceres. Although the presence of ammoniated material on Ceres is generally interpreted as a signature of an outer Solar System origin \citep[e.g.,][]{de2015ammoniated}, our results suggest that Ceres could have formed in situ and accreted volatiles locally. In contrast, ammoniated material on intermediate-sized ($\simeq 100$--$200~\mathrm{km}$) MBAs, if confirmed, does serve as the record of the dynamic history of the early Solar System. Accordingly, we suggest that future asteroid observations should prioritize investigating 100--200 km asteroids that exhibit 3.1~$\mu$m absorption feature, as they provide key constraints on the formation of the asteroid belt.

We predict that 100--200 km ammonia-bearing asteroids may share compositional and isotopic similarities with comets, supporting their origin from distant reservoirs. The isotopic composition of samples from these ammonia-bearing asteroids, if obtained, may differ from that of carbonaceous chondrites and previous sample-return missions, such as those conducted by Hayabusa2 \citep{2022SciA....8D8141H} and OSIRIS-REx \citep{2025NatAs...9..199G}, which targeted rubble-pile asteroids reassembled from the fragments of larger parent bodies. Conversely, if future returned samples from ammonia-bearing asteroids exhibit isotopic compositions indistinguishable from those of previously sampled C-complex asteroids—despite their distinct spectral classifications (e.g., 3 $\mu$m band types) and CCs, this may suggest that these bodies share a common origin in the outer Solar System and were subsequently implanted into the main belt. Such a finding would support the idea that the volatile-rich population in the asteroid belt primarily comprises objects that migrated inward, rather than acquiring volatiles locally via icy pebble accretion. However, given the current limitations in sampling coverage, caution is needed in generalizing the origins of all volatile-rich asteroids. Furthermore, it is important to note that typical ammonia-bearing asteroids are approximately 200~km in diameter, and such small bodies may have insufficient gravity to support water-rock differentiation, making it difficult to create environments suitable for aqueous alteration to form ammoniated phyllosilicates. For instance, ammoniated saponite requires water-rich conditions with a water-to-rock mass ratio of $\gtrsim 4$ \citep[e.g.,][]{kurokawa2022distant,2024GeCoA.374..264S}, while the incorporation of ammonia into hydrated phyllosilicates, such as montmorillonite, requires a highly concentrated ammonia solution (30 wt\% ammonia solution in water) with a mineral-to-solution ratio of 1:10 \citep{2021NatCo..12.2690S}. 

However, NIR spectroscopic observations of such bodies remain limited, and the interpretation of the weak 3.1~$\mu$m absorption feature is still subject to significant debate \citep{2010AAS...21640905H,2011A&A...526A..85B,usui2019akari,rivkin2022nature}. Ground-based telescopes face inherent challenges in accurately characterizing the 3~$\mu$m band due to atmospheric absorption. Moreover, due to the absence of the 3.1~$\mu$m absorption in carbonaceous chondrites collected on Earth, the specific nature and mineralogical form of the ammonia-bearing carriers on large MBAs remain uncertain. Future work should focus on low-temperature laboratory experiments and thermochemical modeling to investigate the formation and transformation of ammonia-bearing materials under conditions relevant to icy asteroids. Such studies are essential to clarify the origin of the spectroscopic signatures of ammoniated minerals and to better constrain the initial physical and chemical conditions of these bodies—such as the water-to-rock ratio, volatile content, internal structure, and their (post-)aqueous alteration history. In particular, it is important to evaluate whether trace amounts of ammonia—potentially delivered via icy pebble accretion, could suffice to generate ammonium-bearing minerals on asteroid surfaces. Future asteroid surveys should aim for a more extensive investigation of the 3~$\mu$m band across the main belt, particularly by utilizing space telescopes to confirm whether the observed spectral signatures indeed correspond to ammonia-bearing materials.

\subsection{Model limitations}
\label{D4}
Our analysis deliberately isolates the two most influential hydrodynamic parameters—Stokes number (${\rm St}$) of pebbles and turbulent strength ($\alpha$) of gas disk—to evaluate the ability of asteroids to accrete pebbles with minimal degeneracy. Although our study provides useful insights into the icy pebble accretion scenario for large volatile-rich asteroid formation, it is not without limitations. Below, we summarize the main simplifications and discuss their impacts.

\subsubsection{Effect of Disk Thermal Structure and Transient Dust Pile‐up Around Snowlines}\label{3.1}

\begin{figure}[htbp] 
    \centering
    \includegraphics[width=0.4\textwidth]{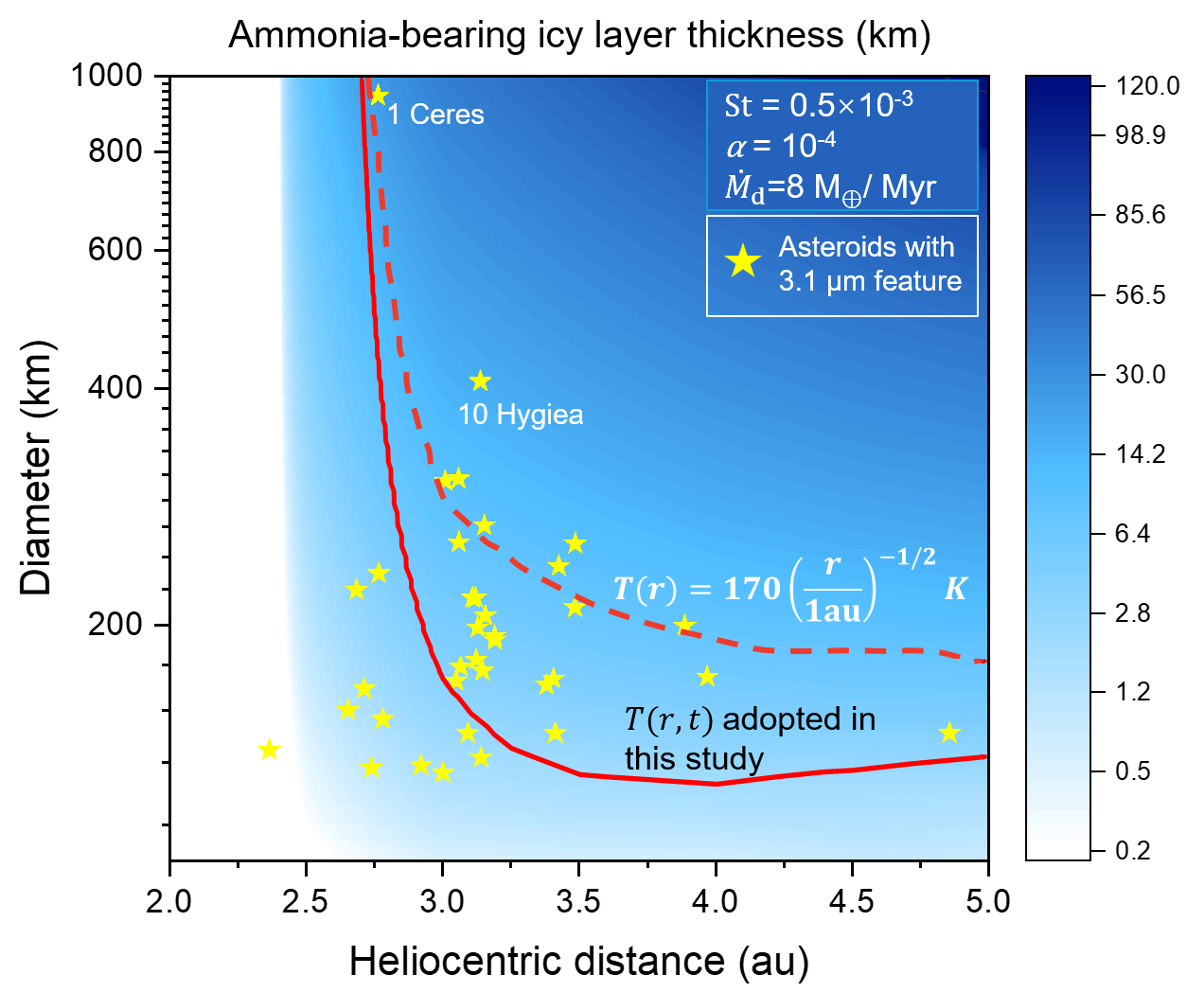} 
    \caption{
    Ammonia-bearing icy layer thickness $\Delta H_\mathrm{NH_3}$ as a function of heliocentric distance and asteroid size, compared with existing large MBAs, assuming a fixed disk temperature profile $T(r)=170\,(r/1\,\mathrm{au})^{-1/2}$ and prescribed snowline migration scenario \citep{sato2016}. Red lines show the boundaries above which $\Delta H$ meets compositional constraints. Dashed line shows for current case ($\mathrm{St}=0.5\times10^{-3}$, $\alpha=10^{-4}$, $\dot{M}_{\rm d}=8~M_\oplus\,\mathrm{Myr^{-1}}$), while the solid line marks the boundary obtained using the time-dependent temperature structure adopted in this work (($\mathrm{St}=0.5\times10^{-3}$, $\alpha=10^{-4}$, $\dot{M}_{\rm d}=7.7~M_\oplus\,\mathrm{Myr^{-1}}$)). The comparison illustrates that disk cooling shifts the boundary outward and allows thicker ammonia-bearing ice layers to be accumulated, particularly for intermediate-sized asteroids.
    }
    \label{compare} 
\end{figure}

The gradual thermal evolution of the protoplanetary disk modifies the radial temperature and pressure gradients, and hence the local pressure gradient $\eta$, the headwind velocity $v_{\mathrm{hw}}$, and the critical Stokes number for efficient pebble accretion. Our model calculates $\eta$ from $T(r,t)$ and fixed gas surface density, $\Sigma_\mathrm{g}$, Figure~\ref{compare} compares our model with a fixed-temperature profile (fixed $\eta$) coupled to a prescribed snowline-migration track \citep[following the method of][]{sato2016}, and shows that disk cooling shifts the efficient-accretion boundary outward and yields somewhat thicker icy layers, especially for intermediate-sized asteroids. Our adopted temperature structure should therefore be regarded as schematic: more comprehensive disk-structure models that self-consistently couple viscous heating, irradiation, and opacity evolution may quantitatively change $\eta$ and the critical Stokes numbers \citep[e.g.,][]{ida2019A&A...624A..28I}, but are unlikely to remove the basic difficulty of building sufficiently thick layers on 100–200 km bodies.

We neglect vapor recondensation feedback which can enhance the surface density of icy dust near the migrating water snowline \citep{Schoonenberg2017,wang2025}. This process may temporarily supply more icy pebbles to asteroids and the phase change process may also change the local pressure gradient $\eta$ which may effect our results \citep{wang2025}. However, as our model calibrates the pebble radial flux $\dot{M}_{\mathrm{d}}$ using Ceres as an upper-limit benchmark, any such enhancement of dust density would only reduce the effective $\dot{M}_{\mathrm{d}}$ required to achieve the observed mass and volatile content of Ceres. Thus, including this effect would not increase the volatile content of 100–200 km-sized asteroids under Ceres formation conditions. For lower-temperature volatiles, ammonia and carbon dioxide account for only 1\% and 20\% of the volatile mass in pebbles, respectively. Their snowlines drift inward to the outer main belt (2.5--3.5~au) only during the late stages of the assumed accretion period (i.e., $\gtrsim 2$~Myr). A recent study suggests that these snowlines may also generate dust pile-ups \citep{2019sf2a.conf..319V}. We acknowledge that incorporating these mechanisms could modestly aid volatile accretion. In contrast, even without considering the local dust pile-up associated with the passage of the water snowline, the accretion of water ice is sufficient for all large asteroids in this study.

\subsubsection{Single‐size Pebble Approximation}\label{3.2}

Our model omits the treatment of a full pebble size distribution and adopts a single size distribution. We compute the accretion rate using the fitting formula of Equations~\eqref{accretion1} and ~\eqref{collisional_efficiency}, which contains an exponential cut-off to bridge the ballistic and headwind regime \citep{visser2016growth}, crate a sharp transition across the critical Stokes number. In reality, pebbles should have a size distribution, and it implies that the effective accretion rate should be interpreted as a mass-weighted average over ${\rm St}$ with a given size distribution, and the transition should be smoothed. Because of this, our single-size approximation may lead to underestimation of the accretion efficiency, especially in the regime of large Stokes numbers ($10^{-3}$--$1$). Recent studies have shown that a full-size distribution can boost 3D headwind accretion by \mbox{$\sim$1–2} orders of magnitude \citep[e.g.,][]{lyra2023}. 

However, the maximum dust size in protoplanetary disks is set by whichever barrier is more restrictive—fragmentation or radial drift \citep{birnstiel2012modeling,cridland2017drift}. In the case of fragmentation-limited regime, laboratory and theoretical studies have shown that when collision speeds approach the fragmentation threshold, dust growth stalls near a characteristic size, and most of the solid mass is concentrated around this peak \citep{Brauer2008,2011A&A...525A..11B,birnstiel2012modeling,Drążkowska2021}. In the drift-limited regime, modeling studies show the Stokes number remains nearly constant for dust grains whose growth is limited by radial drift \citep[drift-limited state,][]{ida2016,taki2021new}, which justifies using a fixed Stokes number as a free parameter in our model. Moreover, previous modeling studies have also shown that results from a single-size approximation closely match those obtained using a full-size distribution for both the fragmentation-limited and drift-limited regimes \citep{birnstiel2012modeling,okuzumi2012rapid,sato2016}. Moreover, the enhancement of the accretion efficiency is negligible for the $\sim120$~km asteroids we studied \citep[with $M \approx 10^{-6}\,M_\oplus$ even under our high asteroid density assumption,][]{lyra2023}. Because we calibrate $\dot{M}_{\rm d}$ using the requirement that Ceres forms, any boost in Ceres' accretion efficiency would reduce the inferred pebble flux, which would not increase (and may further reduce) the volatile delivery to smaller asteroids under our Ceres-forming normalization. Quantifying these distribution effects requires coupling pebble accretion to a dust-evolution and turbulence model, which we leave for future work.

\subsubsection{Effect of Turbulence on Pebble Motion}\label{3.3}
Our model omits the effects of turbulence on pebble random motions. In reality, turbulence excites random motions of pebbles, so that the encounter speed $\Delta v$ between pebbles and planetesimals fluctuates around the $v_{\mathrm{hw}}$. This would be analogous to considering the full size distribution, thereby increasing the probability of pebble accretion for a given St and broadening the transition range between the headwind and ballistic regimes. However, previous studies indicate that the effect that increases the encounter speed between pebbles and planetesimals and suppresses 3D headwind accretion \citep{OL18}, is most severe in the outer disk (at $\sim$~30\,au) and is considerably weaker in the region we focus on (2--5\,au). We therefore consider neglecting the the effect of turbulence to be a reasonable and sufficient first-order approximation for exploring the impact of turbulence on pebble accretion in the main belt region. We also neglect the modification of the radial drift velocity $v_\mathrm{r}$ when $\mathrm{St}/\alpha \sim 1$, where particle advection with the gas accretion flow and turbulent diffusion can contribute to the net radial flux and change the drift speed by a factor of $\sim 2$ \citep{taki2024}.

\subsubsection{Gas Disk Accretion}\label{3.4}
In reality, gas accretion will govern the temperature of the protoplanetary disk, whereas our model directly imposes a given temperature distribution. Besides, our model neglects the effect of gas inflow on dust motion, which may affect the radial drift of small dust that is tightly coupled with gas. Thus, a static gas disk might overestimate pebble accretion. However, \citet{taki2024} shows that typical viscous gas accretion flows modify dust drift speeds only by factors of order unity, particularly when the turbulence parameter $\alpha$ is comparable to or smaller than the particle Stokes number. Thus, while a static gas disk might slightly overestimate pebble accretion, the overall effect is limited and does not significantly impact our conclusions.

\subsubsection{Formation Timing of Planetesimals}\label{3.5}

Our model assumes that all planetesimals are already present at \(t=0~\mathrm{Myr}\). We further assume that the pebble radial flux, \(\dot{M}_{\mathrm d}\), is steady over the disk lifetime (\(\sim 3~\mathrm{Myr}\)), with its magnitude constrained by requiring Ceres to grow from a rocky embryo. In reality, planetesimal formation may have occurred over an extended interval, leading to differing timing for starting the pebble accretion \citep{johansen2021pebble}, which could modify the amount of volatile material acquired via pebble accretion.

The formation timing of the primordial asteroids in the main belt may be later than assumed in this study, or may differ between individual bodies. Streaming instability is a candidate mechanism for planetesimal formation \citep{Youdin2005}. Several competing effects may operate during disk evolution. In smooth disk-evolution scenarios without long-lived pressure bumps that halt radial drift, and with dust coagulation, radial drift tends to reduce both the dust surface density and the characteristic particle size as the outer disk becomes depleted of solids \citep{lambrechts2014forming,sato2016}. As a result, the midplane dust-to-gas ratio and the Stokes number decrease, thereby narrowing the parameter space for triggering streaming instability at late times \citep{2015A&A...579A..43C}. Conversely, as gas is dispersed, the dust-to-gas ratio and the Stokes number of solids tend to increase \citep[e.g.,][]{okuzumi2025}, which can facilitate stronger particle concentration and potentially shift the characteristic planetesimal size toward larger values. Geochemical constraints from meteorites and thermal-evolution models of planetesimals provide an additional perspective \citep{Alexander2018,Lichtenberg2021}; earlier-formed planetesimals (\(\gtrsim 30~\mathrm{km}\)) are strongly heated by the decay of short-lived radionuclides and thus lose their volatiles. Motivated by these constraints, volatile-rich large MBAs may be more consistent with comparatively late formation \citep[e.g., \(t\gtrsim 0.7~\mathrm{Myr}\);][]{Lichtenberg2021}, because earlier growth to large sizes would raise peak interior temperatures and prolong thermal processing, decreasing the survival probability of any subsequently accreted icy layer \citep{takir2023late,2025SciA...11.3283C}.

Within our model, if the entire rocky planetesimal population forms systematically later than $t=0$, the effect is largely equivalent to adopting a colder disk model, and the overall trends in volatile acquisition remain similar, as shown in Figure \ref{compare}. Relative formation sequences between the Ceres embryo and the intermediate-sized asteroids can modify the amount of volatiles they acquire. If the Ceres embryo forms later than the intermediate-sized asteroids, those earlier-formed planetesimals would have more time to accrete volatile-rich pebbles and could end up with higher volatile fractions. For instance, if the intermediate-sized asteroids form at time \(t_{\rm form}=t_1\) and the Ceres embryo forms at \(t_{\rm form}=t_2\) (with \(t_1<t_2\)), then the earlier-formed bodies have a longer time interval available for pebble accretion, \(\Delta t = t_{\rm end}-t_{\rm form}\), where we take \(t_{\rm end}=3~\mathrm{Myr}\). In our fixed-\(\dot{M}_{\rm d}\) framework, the accreted volatile mass (and hence the thickness of the accreted icy layer) approximately scales with \(\Delta t\). Therefore, the total volatile mass delivered to a specific body scales as \(M_{\rm vol}(t_1)/M_{\rm vol}(t_2)\approx \Delta t_1/\Delta t_2\), where \(\Delta t_i\equiv t_{\rm end}-t_i\). This may affect our conclusions, allowing intermediate-sized asteroids to actually experience accretion of more volatile components with relatively large Stokes numbers ($\text{St} \gtrsim 10^{-3}$). However, if the snowline has already migrated past the formation region, planetesimals could be volatile-rich from birth rather than through subsequent icy pebble accretion. A late-forming Ceres embryo would then be able to match its inferred volatile fraction without requiring the same pebble-flux constraint. A detailed analysis of this point is left for future studies.

\section{Summary} \label{sec:summary}

The main asteroid belt is a relic of the early Solar System, preserving crucial evidence of planetary formation and volatile delivery. It exhibits a taxonomic gradient with heliocentric distance (from anhydrous to volatile-rich surfaces). Specifically, volatile-rich asteroids may originate from implantation of outer Solar System bodies or from in-situ accretion of inward-drifting icy pebbles. In this study, we investigated whether pebble accretion, coupled with snowline migration, could account for the observed volatile distribution among MBAs under an early Solar System scenario in which their embryos formed in situ. Using a simplified disk model, we explored how pebble flux, Stokes number, and turbulence influence the accretion of icy pebbles onto rocky planetesimals as the snowline migrated inward.

We calculated the growth and volatile acquisition of MBAs with a moderate pebble radial flux, one consistent with Ceres's growth from its rocky embryo. We compared the accretion outcomes with observational constraints. The results show the feasibility of water and ammonia ice accretion, and highlight challenges in supplying volatiles to bodies with $D< 200\ \mathrm{km}$, which require smaller pebbles.

Our results indicate that large MBAs, such as Ceres, could have acquired sufficient volatiles through icy pebble accretion under favorable conditions. Moreover, even under relaxed conditions for pebble accretion onto 100--200 km-sized planetesimals, the delivery of ammonia ice to them needs pebbles with small Stokes numbers ($\text{St} \sim 10^{-4}$). If such conditions cannot be achieved in the disk, this may imply that such MBAs may have formed in the outer Solar System beyond the ammonia snowline and were later implanted into the main belt.

Our findings underscore the importance of telescopic observations and spacecraft explorations for 100--200 km MBAs, as they serve as key tracers of volatile transport mechanisms in the early Solar System. In addition to expanding surveys of asteroids to obtain more 3~$\mu$m band spectral information and samples, future research should also test the feasibility of sustaining small pebbles ($\text{St} \sim 10^{-4}$) in realistic disk environments. Furthermore, theoretical and experimental studies should focus on simulating aqueous alteration processes. This will help bridge the gap between protoplanetary disk and planet formation models and the complex mineralogical record preserved in meteorites and asteroids.

\begin{acknowledgments}
We are very grateful to Dr. Tomokatsu Morota and Dr. Kosuke Kurosawa for providing the knowledge needed to estimate the impact crater excavation. This work was supported by JST SPRING Grant Number JPMJSP2108 and JSPS KAKENHI Grant Number 25KJ1007, 20KK0080, 22H01290, 22H05150, 21H04514, 21K13983, and 23K22561. The authors are deeply grateful to the anonymous reviewer for their insightful and constructive comments, which substantially reshaped and strengthened the core of this study.

\end{acknowledgments}

\bibliographystyle{aasjournal}  
\bibliography{references}  

\end{document}